Article

# Algorithmic Decision-Making in AVs: Understanding Ethical and Technical Concerns for Smart Cities

Hazel Si Min Lim and Araz Taeihagh *

Lee Kuan Yew School of Public Policy, National University of Singapore, 469B Bukit Timah Road, Li Ka Shing Building, Singapore 259771, Singapore; a0129822@u.nus.edu
* Correspondence: spparaz@nus.edu.sg; Tel.: +65-6601-5254

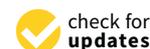

**Abstract:** Autonomous Vehicles (AVs) are increasingly embraced around the world to advance smart mobility and more broadly, smart, and sustainable cities. Algorithms form the basis of decision-making in AVs, allowing them to perform driving tasks autonomously, efficiently, and more safely than human drivers and offering various economic, social, and environmental benefits. However, algorithmic decision-making in AVs can also introduce new issues that create new safety risks and perpetuate discrimination. We identify bias, ethics, and perverse incentives as key ethical issues in the AV algorithms' decision-making that can create new safety risks and discriminatory outcomes. Technical issues in the AVs' perception, decision-making and control algorithms, limitations of existing AV testing and verification methods, and cybersecurity vulnerabilities can also undermine the performance of the AV system. This article investigates the ethical and technical concerns surrounding algorithmic decision-making in AVs by exploring how driving decisions can perpetuate discrimination and create new safety risks for the public. We discuss steps taken to address these issues, highlight the existing research gaps and the need to mitigate these issues through the design of AV's algorithms and of policies and regulations to fully realise AVs' benefits for smart and sustainable cities.

## 1. Introduction

Smart and sustainable cities have been increasingly emphasised around the world to resolve the challenges associated with rapid urbanisation and population growth [1–4]. Amid the proliferating initiatives to develop smart cities, conceptions of the latter have emerged as a point of contention among scholars, who have pointed out the heavy emphasis of smart city concepts on technologically driven solutions and economic imperatives and the lack of evidence demonstrating their alignment with environmental and social sustainability [4–7]. In the transportation industry, these initiatives take the form of smart mobility solutions ranging from shared mobility services, electric vehicles and bikes, autonomous vehicles (AVs) and the integration of multiple transportation modes such as rail and cycling to attain safer, more efficient and environmentally sustainable transportation [8–13]. There have, however, been increasing calls for these solutions to not only be adopted for their technological "smartness", but also to be supplemented with more integrated transport planning that serves the needs of people and sustainable mobility objectives [5,6,14].

Among the various smart mobility solutions proposed, AVs offer improved safety, congestion, traffic efficiency, mobility for the elderly and disabled and reduced environmental impact [1,15–18] and many governments around the world have been accelerating AV developments and trials [19,20]. Underpinning the technological "smartness" of AVs are connected sensors that retrieve data on its surroundings and algorithms that process the data, make, and execute driving decisions through the vehicle's actuators. The quality of data retrieved by the AV's sensors is critical for decision-making [21], and the efficiency, precision and reliability of decision-making algorithms allows AVs to surpass the typical human driver in performing driving tasks [22]. A major component in AVs is machine-learning (ML) [23] algorithms that continuously learn and adapt to new information, which is essential for responding to unexpected situations and providing on-demand transportation services [24]. Algorithms' independence from human input and ML's data-driven nature allows AVs to significantly reduce or eliminate human errors that have been responsible for 90% of road fatalities, such as speeding, alcohol impairment, distractions and induced fear [25,26].

However, an overemphasis on technological solutions alone for economic development could risk neglecting social and environmental considerations and thus hinder true "smartness" [27]. Scholars have cautioned against rushing to develop smart mobility solutions such as AVs without being prepared to manage their potential "negative externalities" [1,28]. In particular, issues in algorithmic decision-making in AVs can have undesirable effects on safety and equity. Firstly, data mining processes in AVs are susceptible to biases that lead algorithms to prioritise the safety of certain groups of road users over others and thus, perpetuate discrimination [29,30]. Secondly, many scholars stress the need to design algorithms with ethical considerations to ensure that AVs make ethical driving decisions [31,32], but the proposed approaches entail various limitations that can lead to unsafe or unfair driving decisions [33,34]. Thirdly, stakeholders in the AV ecosystem can be motivated by perverse incentives to design algorithms that discriminate between different groups of road users and that promote risky driving behaviour [31,35]. Fourthly, technical issues in the AV system can undermine the accuracy of algorithmic decisions and create new safety hazards [36,37].

These new safety risks and discrimination that can potentially arise from the ethical and technical issues in AVs' algorithms can undermine public well-being and social equity, and thus are obstacles to real "smartness" and sustainability, but limited research exists to assess these issues. Furthermore, holding stakeholders accountable for AVs' algorithms are critical to mitigate these negative consequences, but exploration of such accountability mechanisms remain nascent in the literature [17,31]. Transparency has also been criticised as insufficient to explain highly complex algorithms and potentially inhibitive for innovation due to excessive "regulatory constraints" on algorithm designs [38]. New solutions to account for and mitigate these new safety risks and discrimination are required for AVs to be considered a truly smart and sustainable transportation solution. Thus, we seek to address the following questions:

(a) What are the issues in algorithmic decision-making in AVs?

(b) How do these issues create concerns for safety and discrimination and impede the move towards smart and sustainable cities?

(c) What are the proposed solutions and steps taken to address these issues?

Section 2 provides the background information about AVs and algorithms, followed by Section 3, which describes our research methodology. In Section 4, we discuss the importance of algorithmic decision-making in AVs for smart and sustainable cities and their emerging issues. Sections 5 and 6 explore the ethical and technical issues in algorithmic decision-making in AVs and the proposed solutions to address these issues. We then discuss our findings, the implications for smart and sustainable cities, and the research gaps to be addressed in Sections 7 and 8.

## 2. Background

### 2.1. AVs

The Society of Automotive Engineers (SAE) [39] classifies vehicles according to different levels of autonomy. At Levels 1 and 2, the human driver performs the main driving operations, but the vehicle is aided with advanced driver assistance systems, such as rear-view video systems and automatic emergency braking. At Level 3 (conditional automation), the vehicle can perform all dynamic driving tasks such as steering, acceleration, and monitoring the environment, but the human driver is required to resume control occasionally [39,40]. Vehicles classified under Levels 4 and 5 of autonomy are considered highly and fully autonomous respectively as they can engage in all the driving tasks without human intervention [41].

The implementation of vehicles at Levels 4 and 5 is now possible due to rapid technological advancements in hardware and software systems—sensor-fusion technology and computer vision allow the vehicle to detect, trace and manoeuvre safely around obstacles and under a wide range of environmental conditions [42,43]; High-performance computing enables the vehicle to process vast amounts of data to understand its environment and make spontaneous driving decisions; and communication technologies enable the vehicle to exchange information with and learn from other vehicles [44,45]. Given the accelerating technological developments and roll-out of AV trials, AV technology is expected to advance rapidly and surpass human capabilities over time [38]. Throughout this study, we focus on vehicles classified under SAE's Levels 4 and 5 of autonomy, which we will refer to as "AVs".

*2.2. Algorithms*

Algorithms form the mechanisms for decision-making in Artificial intelligence (AI), which was developed with the intention of mimicking human intelligence. This study refers to the notion of weak AI, which is defined in terms of the "specific tasks that require single human capabilities, e.g., visual perception, understanding context, probabilistic reasoning and dealing with complexity". Strong AI implies "a system with human or superhuman intelligence" to perform human thinking such as ethical judgements, "symbolic reasoning" and "managing social situations" [8]. AI may be conceived of firstly as traditional computer programs that rely on rule-based algorithms to classify information [46,47] and secondly as a system that can solve problems using high-level reasoning and under uncertainty, and learn from experience [38]. AI under the second conception uses ML [23] algorithms that do not merely process data through "static mathematical models" but also learn from historical data to make future decisions [48]. ML algorithms learn by processing input and output data to model their understanding of the world, typically by assigning weights to different input variables that represent their relative importance in determining the outputs. The internal model is then optimised against a new set of data to ensure its predictive accuracy, where the ML algorithm selects the model that best aligns with the real-world by minimising its prediction error according to some pre-defined preferences and decision-making criterion [38].

In AVs, algorithms are embedded in hardware and software. The AV's hardware comprises sensors to obtain data about its environment [49], vehicle-to-vehicle and vehicle-to-infrastructure communication technologies to exchange information with other connected vehicles and infrastructure, and actuators (e.g., steering wheel, brakes, simulators) to execute the AV's physical movements [37,50]. These hardware components interact with the AV's software, which comprises perception, decision-making (or planning) and control. Perception refers to the collection of information and knowledge from the environment by the AV system through its sensors and communication networks; Decision-making enables the AV to meet its goals through processes including mission planning that involves decision-making to meet "high-level objectives" such as deciding which route to take, behavioural planning that involves producing "local objectives" such as switching lanes and overtaking, and motion (or local) planning that produces the steps required to achieve local objectives, such as reaching a target destination [37], whereby algorithms make decisions based on selected preferences and criterion; and control algorithms execute these decisions by calculating the inputs for the AV's actuators, such as the steering angle and vehicle speed, for the AV to follow a given trajectory [50,51]. The proper functioning of algorithms in all software and hardware components is critical for the AVs' safe operation.

## 3. Methodology

The methodology of this study was designed around our three research questions. We first identified the issues arising from algorithmic decision-making in AVs that were most saliently discussed in the AI and ML, robotics, transportation, smart and sustainable cities, philosophical, policy, planning, and legal literature. We used the key words associated with algorithms, AVs and smart and/or sustainable cities (see Table 1) in combination with "issue(s)", "impact(s)", "consequence(s)", "effect(s)", "limitation(s)", "concern(s)", "implication(s)", "risk(s)". Our search revealed that bias, ethics, and perverse incentives were among the most prominent concerns being raised in the literature regarding algorithmic decision-making in AVs that have significant implications for safety and discrimination. Search results from the technical literature revealed that various aspects of algorithmic decision-making with impacts on AV safety and performance stem from the AV's different components mainly involved in data processing and analysis (i.e., perception, decision-making, control), the computational capacity of existing hardware (e.g., graphical processor units) and testing and verification procedures. For conceptual simplicity, we broadly categorised bias, ethics, and perverse incentives as "ethical issues" and the issues arising from the AV system's perception, decision-making, control and testing and verification as "technical issues". Rather than aiming to construct a fully exhaustive taxonomy of all the possible issues arising from the topic at hand, this categorisation serves as a first step in distinguishing between issues in AVs' algorithmic decision-making that arise from the AV system's technological landscape (i.e., technical issues), with those issues that involve a broader range of actors and ethical implications beyond the technological landscape in the wider AV ecosystem (i.e., ethical issues).

**Table 1.** Keywords used to identify articles relating to algorithmic decision-making in AVs and smart and sustainable cities.

| Topic | Keywords |
|---|---|
| Algorithm | algorithm(s), algorithmic, algorithmic process(es), algorithmic decision-making, artificial intelligence, machine-learning, machine-learning algorithm(s), autonomous system(s), autonomous decision-making |
| AVs | autonomous vehicle(s), driverless, driverless vehicle(s), self-driving vehicle(s), unmanned autonomous vehicle(s), autonomous car(s), driverless car(s), autonomous vehicle technology |
| Smart and/or sustainable cities | smart cities, smartness, sustainable cities, sustainability, smart and sustainable, smart mobility, smart transportation, intelligent transportation, sustainable mobility, sustainable transportation, smart technology, sustainable technology |

After identifying the set of issues arising from algorithmic decision-making in AVs, we conducted an in-depth analysis of each issue, its implications for safety and discrimination, and its significance for smart and sustainable cities. This involved searching the keywords in Table 1 in conjunction with the following keywords in Table 2. Lastly, we identified the proposed solutions and/or steps taken to address these issues by searching these terms in combination with the terms relating to solutions and governmental action, which includes "solution(s)", "government", "governance", "policy", "regulation(s)", "guideline(s)", "law(s)", "proposal(s)" and "proposed solution(s)".

Our search covers research articles dated from 2000 onwards and was conducted primarily on established academic databases, namely Scopus, Web of Science, ScienceDirect and Springer, as well as on Google Scholar when the results from database searches were insufficient. We utilised mainly peer-reviewed journal articles to analyse the issues from algorithmic decision-making in AVs and supplemented the analysis on technical issues with conference proceedings pertaining to AI, ML and engineering, such as those published by the Institute of Electrical and Electronics Engineers (IEEE) and Association for the Advancement of Artificial Intelligence (AAAI). We also included government reports and policy documents to explore the governmental actions taken to address these issues.

Table 2. Keywords used to identify articles about the issues arising from algorithmic decision-making in AVs and their implications for safety and discrimination.

| Issue | Keywords |
| --- | --- |
| Bias and/or discrimination | bias(es), data bias(es), biased, discrimination, discriminatory, discriminate, disparate outcome/treatment/effect, differential outcome/treatment/effect, disparity, unequal, (in)equality, (in)equity, (un)fair, (un)fairness, prejudice |
| Ethics | Ethic(s), (un)ethical, moral(s), value(s), ethical standard(s)/value(s)/principle(s)/rule(s), societal standard(s)/value(s)/principle(s), ethical/moral dilemma(s), ethical/moral programming, machine ethics, moral agency, roboethics, ethical theory, thought experiment(s), trolley problem(s)/scenario(s)/case(s), ethical/moral judgement(s), risk allocation |
| Perverse incentives | Incentive(s), profit(s), profitable, economic incentive(s), commercial incentive(s), motivation(s), stakeholder(s), manufacturer(s), programmer(s), designer(s), user(s), passenger(s), customer(s), consumer(s) |
| Technical | Technical, technology, technological, technical/technological limitation(s), system, software, hardware, system component(s), design, programmer, code, operation, operating, operator(s), maintenance, malfunction(s), error(s) |
| Perception | Perception, sensing, sensor(s), camera, visual, vision, machine perception, machine interpretation, object/obstacle/image detection, image recognition, perception system, environmental perception, driving/road environment |
| Decision-making | Decision-making, planning, path/motion/local/trajectory/route planning, path/trajectory generation, optimisation, modelling, logic, human-machine interaction, human-machine interface, uncertainty, decision-making rule(s)/criteria/criterion/preference(s) |
| Control | Control, controller, control system, vehicle/vehicular control, path/trajectory tracking, vehicle motion, control techniques, control strategies |
| Testing and verification | Test(s), testing, trial(s), verification, verify, validate, validation, requirement(s), verification method(s)/technique(s)/tool(s), validation method(s)/technique(s)/tool(s), assessment |
| Safety | Safety, safe, accident(s), risk(s), collision(s), collision avoidance, fatalities, injury, injuries, harm(s) |

## 4. Algorithmic Decision-Making in AVs for Smart and Sustainable Cities

The notion of sustainable cities emerged in the 1990s [52] and while various definitions of the term exist, sustainability may be understood as a state in which a city's social, economic and environmental needs can be met while ensuring the overall system's self-sufficiency and continuous improvement [45,53]. Thus, sustainability entails the simultaneous attainment of economic sustainability through increased efficiency and economic growth, social sustainability through improved equity and public well-being, and environmental sustainability through the smart use of resources [53,54]. Since 2010, interest has significantly shifted towards smart cities, whereby innovative information and communication technologies (ICTs) are utilised to improve the economic competitiveness, standard of living and operational efficiency of cities [21,55]. This is driven by the use of connected devices such as sensors, actuators and wearables that are able to store and transmit data through the Internet, i.e., the "Internet of Things" (IoT), which forms the backbone of smart infrastructure in smart cities as these devices can interact and synchronise their actions across multiple smart applications such as community development, grid distribution and transportation [4,56]. The concept of smart cities goes beyond earlier concepts of "information city", "digital city", "sustainable city" and "intelligent city" by emphasising the use of technology to serve the needs of people [52,55]. Scholars argue that a city can only be considered truly smart when technological solutions are utilised not only for greater economic efficiency but also social and environmental sustainability [7].

A major component of smart cities is the smart mobility agenda that aims to integrate IoT and smart vehicle technologies into the transportation system, among which AVs play a central role [57]. Smart mobility integrates intelligent transportation systems, which comprise of connected road infrastructure and smart vehicular technologies such as AVs, to improve transportation outcomes such as safety, access to mobility and traffic management, which are key to enhancing public well-being and economic efficiency [7,10,17,58]. In AVs, data collection conducted by on-board sensors directly influences the AV's situational awareness and the algorithms that make decisions such as route planning, obstacle avoidance and reverting control to humans [56,59]. Secondly, the highly autonomous nature and high computational power of the AV's algorithms significantly increases the efficiency, accuracy and timeliness of decision-making relative to human drivers [24]. ML capabilities enable the AV system to continuously improve and adapt its decision-making processes to environmental changes over time, which are critical to maintaining and improving users' safety and security when faced with manufactured and natural hazards [24,60]. Coupled with external connectivity from the use of IoT devices, ML capabilities also enable AVs to provide and tailor on-demand transportation services to changing consumer demands [24]. Thus, the "smartness" and sustainability of AVs hinges upon the embedded connected devices that store and transmit data as well as decision-making algorithms that make "precise and real-time decisions" [21] as these processes directly influence road safety, traffic efficiency, the quality and range of transportation services and other purported benefits for sustainability [24,61].

Despite the hype over AVs, multiple ethical and technical issues have surfaced regarding algorithmic decision-making that can undermine AVs' promise as a truly smart and sustainable transportation solution. Firstly, algorithms are susceptible to biases originating from the data and the human designer [62,63]. Based on mathematical correlations learnt from the data, algorithms can penalise certain personal characteristics [64], such as a pedestrian's gender, age or physical health, and the associated individuals, resulting in the AV allocating more risks to certain groups of individuals over others and thus, creating discriminatory driving outcomes [31]. Secondly, stakeholders in the AV value-chain can design algorithms in ways to maximise profit rather than for public safety, particularly as sustainability is "not central" to the developments of connected vehicles and AVs, which is instead motivated by technological development for its own sake and/or its profitability [5]. Thirdly, AVs' allocation of risks is inherently an ethical decision that necessitates ethical rules to inform algorithmic decision-making, but the choice of rules and approach to program these rules can yield new safety risks and discriminatory risk allocations [65]. Lastly, algorithms in the AV's software and hardware components contain many unresolved limitations that can undermine the reliability, safety of and public's trust in AVs. These issues are more difficult to recognise and correct in ML than rule-based algorithms as the former model logics that are not explicitly programmed and are not easily understood by humans [47]. Unlike rule-based algorithms where the logics of decision-making are explicitly specified in advance by the programmer, ML algorithms learn from the data by constructing "highly nonlinear correlations" between inputs and outputs, which cannot be easily understood by humans [47]. These new safety risks and potential discrimination resulting from AVs can undermine social equity and public well-being that can dampen consumer acceptance of AVs, which is vital for the technology's adoption and realising its benefits for smart and sustainable cities [66]. In the next two Sections, we examine the ethical and technical concerns in algorithmic decision-making in AVs, their implications for safety and discrimination, and discuss the steps taken to address these issues.

## 5. Ethical Concerns from Algorithmic Decision-Making in AVs

This Section explores ethical issues associated with algorithmic decision-making in AVs, their implications for AV safety risks and discrimination and the steps taken to tackle these issues. Section 5.1 discusses the sources of bias in AVs' algorithms that can yield discrimination by disproportionately allocating more safety risks to some groups of individuals. Next, Section 5.2 explores approaches to incorporate ethics into AV algorithms' decision-making and highlight their implications for AV safety and discrimination. Lastly, Section 5.3 examines how the incentives of AV stakeholders shape AV algorithms' design and resulting decisions that can introduce new safety risks and discrimination.

*5.1. Bias*

A system is considered biased when it contains "intended" or "unintended" characteristics that unfairly discriminate against certain individuals or groups of individuals in society [67]. In American anti-discrimination law, discrimination exists when there is disparate treatment, which is the "discriminatory intent or the formal application of different rules to people of different groups", and/or disparate impact, which is the result that "differ for different groups" [29,68]. Bias can be introduced into AVs during the human designers' construction of the datasets, models, and the parameters of the algorithm, which potentially leads to unfair or discriminatory allocations of safety risks [46,68,69]. Firstly, statistical bias exists when the input data are not statistically representative of the overall population [67]. For instance, training an AV using data from only one country could result in the AV learning localised patterns and not accurately modelling driving behaviours that apply in other countries or contexts [30]. Thus, the under- or overrepresentation of certain groups in the data can lead to inaccurate classifications and biased outcomes [63,70]. Secondly, the algorithm can be biased relative to legal and moral standards if it utilises sensitive input variables [30]. Individual-specific characteristics, such as a person's age and gender that are used as decision-making criteria can be penalised or privileged by the AVs' algorithms to meet the algorithm's pre-defined preferences, such as prioritising the safety of children or minimising the total quantity of harm, causing more safety risks to be allocated to individuals that share the penalised characteristics [31]. These forms of bias can be introduced unintentionally or intentionally by algorithm designers and AV manufacturers to maximise profits, such as prioritising the safety of AV passengers to maximise profits, and this is exacerbated by the lack of legal frameworks to hold these stakeholders accountable [31]. Section 5.2 explores various types of ethical preferences to which AVs may be programmed to follow and their implications of AV safety risks in greater detail, and Section 5.3 explores how perverse incentives influence the choice of preferences that are programmed into AVs' algorithms.

Lessening bias in algorithms is therefore crucial to mitigate discriminatory outcomes from AVs. In autonomous systems in general, scholars have recommended ways to detect and offset the effects of bias, such as modifying algorithmic outputs to balance the effects of bias between protected and unprotected groups, introducing minimally intrusive modification to remove bias from the data [71], incorporating individuals from potentially discriminated groups [72], testing techniques to measure discrimination and identify groups of users significantly affected by bias in software [73,74] and creating algorithms that certify the absence of data bias [75]. Apart from bias originating from the data and selection of variables and criterion, Danks and London [30] recommend clarifying ethical standards such as fairness to evaluate bias. Furthermore, scholars recommend increasing transparency to identify biases [76], such as designing algorithms whose original input variables can be traced throughout the system (i.e., traceability) [72] and auditing algorithms to enhance their interpretability so that biases can be detected and the system's outputs can be verified against safety requirements [70,77].

However, there are challenges in identifying bias in algorithms and their discriminatory effects. Firstly, many algorithms are designed to be highly complex for greater accuracy, but this renders the algorithm opaque and difficult to interpret even by the designers themselves, concealing the sources of bias [63,70]. Secondly, as ML algorithms make decisions mainly based on the training data that changes over time [68], it is difficult to predict potentially discriminatory effects in advance. Humans are also excessively trusting and insufficiently critical of algorithmic decisions due to the popular perception of algorithms as objective and fair, a problem referred to as "automation bias" and the seemingly "objective" correlations that the algorithm learns from the data makes it difficult to legally establish discriminatory intent in algorithms [31,48,64]. An emerging issue is the aggregation of individually biased outcomes when AVs with similar preferences are deployed on a large-scale, as doing so would centralise and replicate algorithmic preferences along with their individually biased risk allocation decisions. This could lead to the same groups of people being consistently allocated more safety risks and perpetuate systemic discrimination, which is more difficult to detect as it results from the accumulation of similar driving outcomes [31,78].

Actions taken thus far to tackle algorithmic bias and discrimination are not specific to AVs, ranging from releasing voluntary AI guidelines, improving the design and testing of algorithms in AI systems, and the EU's General Data Protection Regulation (GDPR). Firstly, governments in Japan and Singapore have released voluntary guidelines for AI that emphasise on explainability and verifiability of AI-driven decisions, fairness to mitigate discrimination, increasing transparency through information disclosure [79] and establishing open communication channels between customers and stakeholders across the value-chain [80]. Singapore's guidelines recommend internal governance practices to increase the accountability of AI-deploying organisations and to mitigate algorithmic discrimination, such as new oversight mechanisms to clarify responsibilities and practices for accountability in operations management and systems design [80]. Secondly, the government in South Korea aims to develop techniques to detect data bias, correct software errors, and test ethical standards at every stage of AI development [81] and the UK government will collaborate with the Alan Turing Institute to develop AI talent and auditing tools to mitigate "social inequalities" resulting from algorithmic decision-making [82]. Lastly, the EU passed the GDPR that prohibits any automated decision that utilises sensitive personal data and that notably affects data subjects in the EU. The GDPR also mandates a right to explanation, which aims to increase the interpretability and transparency of automated decisions by requiring firms to provide data subjects with "meaningful information about the logic involved" in "concise, intelligible and easily accessible" forms [83].

*5.2. Ethics*

Ethics are inherent in various driving scenarios as it involves allocating risks among multiple persons during accidents [18,84] and during routine driving scenarios, such as deciding the following distance from a nearby vehicle. As the AV's risk allocation decisions will be judged both by traffic laws and ethical standards [85,86], scholars highlight the need for AVs to follow ethical rules, which we refer to broadly as ethical theories, principles, norms and values in their decision-making, which can be formalised and designed in AVs using multiple proposed approaches.

Proposals to formalise ethical rules for AVs have revolved around the use of ethical dilemmas in thought experiments [32,84,87]. An ethical dilemma is a situation where it is "impossible to make a decision among various possible decisions without overriding one moral principle" [88]. A popular thought experiment is the trolley problem that can be illustrated through a hypothetical scenario where an AV's brakes are faulty and can either continue in its current path to crash into five pedestrians or swerve to crash into one pedestrian [89]. While trolley problems can reveal individuals' ethical preferences and key decision-making criteria [65,88,90], they make unrealistic assumptions about driving scenarios, such assuming that outcomes are completely certain and that the passenger can choose how harm is distributed [35,86,90,91], are susceptible to inconsistencies in ethical reasoning among its participants [92] and do not consider the effects of aggregating decisions that are ethically justifiable on their own but that potentially create larger systemic patterns, such as discrimination [31]. Other assumptions include the presence of only one decision-maker, the immediacy of the decision and the restriction of considerations [91].

Apart from thought experiments, two broad technical approaches have been proposed to program ethical rules into AVs' algorithms. The "top-down" approach involves mapping a set of ethical theories, such as utilitarianism and deontology, to computational requirements and programming them into the algorithm [93,94], but programming each ethical theory entails their own set of limitations and issues that can undermine AV safety and also perpetuate discrimination.

Utilitarianism emphasises on the morality of outcomes and, in the context of AVs, implies the programming of algorithms to minimise the total quantity of harm from accidents [31,95], but doing so can introduce bias and implementation challenges that can create new safety risks and potential discrimination. Utilitarian algorithms would compute all possible outcomes, alternative actions and their associated consequences, and minimise a cost function (expected total quantity of harm) [65,96]. However, as an optimisation problem that minimises collective rather than individual harms, utilitarian algorithms would not consider equity or fairness and may in fact use inappropriate characteristics as decision-making criteria, leading to biased risk allocation decisions (see Section 5.1). For instance, the algorithm may choose actions that allocate more safety risks to "more protected" road users, such as those wearing a helmet, as they would suffer fewer injuries compared to other "less protected" road users [97]. This could promote "false incentives", such as not wearing a helmet on a motorcycle, which discriminates against those who took safety precautions [89,95]. Secondly, safety risks can emerge from the technical challenges of implementing utilitarian ethics, such as potential inaccuracies and delays in machine perception that undermine the AV's ability to compute all possible outcomes and actions within a short timeframe and the challenges of defining the algorithm's decision-making criteria and cost function [34,41,96].

In contrast, deontology emphasises on actions being motivated by respect for all humans [98], which can be implemented by explicitly programming ethical rules in AVs in a hierarchical manner, but doing so introduces other challenges that could yield new safety hazards for the AV. While deontological rules make explicit the reasoning behind the algorithm's decision-making (e.g., Asimov's Three Laws of Robotics) [36,96], the algorithm may be forced to make sub-optimal decisions to strictly adhere to its rules, such as instructing the AV stop when rules conflict or cannot be satisfied. This creates safety risks for other road users and hinders the AV's adaptability to new circumstances, unlike utilitarian-oriented algorithms that can easily adjust probabilities and magnitudes of outcomes to optimise decisions [65]. Secondly, deontological rules may not cover all kinds of driving scenarios [33,95] and even so, path dependencies can arise if the AV is trained on any particular order of scenarios [36]. Thirdly, many deontological principles are embedded in legal ambiguities of existing traffic laws that cannot be explicitly represented in algorithms, such as the different definitions of "obstruction" or "safe" in different scenarios [33,36]. Given the individual limitations of utilitarianism and deontology, scholars have advocated combining both theories to broaden the AV's perspective of the situation before making a decision [88] and such combinations appear to work well in practice, such as in many organ donation programmes—deontological ethics justifies the first come, first served practice while utilitarian ethics justifies the practice of prioritising the sickest recipients.

The limitations of the top-down approach can be offset by the bottom-up approach, whereby the algorithm constructs its own rules by learning from past driving experiences and human judgements that are deemed morally correct [93,94,99,100], but doing so can also introduce its own safety risks to AVs due to several implementation challenges, the potential for the system to override pre-programmed constraints and greater opacity of decision-making. One application of the bottom-up approach is a greedy-search algorithm that continuously searches for the most ethical solution according to human moral values [101]. Scholars argue that a bottom-up decision-making system can potentially be more ethical than that of "any individual human" by eliminating the latter's unique mistakes and identifying other undiscovered ethical principles [88,99]. However, it is difficult to clearly specify a self-learning system's high-level goals and to prompt the system to expand the domain of possible choices to choose from [101]. In addition, self-learning systems can temporarily shift its goal as part of learning [88], which suggests the potential for an AV to display unethical behaviour so that it can learn from these experiences and make more ethical decisions in future. Lastly, the bottom-up approach can increase the system's opacity as the logic behind the algorithm's self-constructed rules is not known [35,100], exacerbating the challenge of identifying bias, ethically questionable rules, and errors.

Several governments have taken non-regulatory measures to address the ethical issues in algorithmic decision-making that are not AV-specific through voluntary guidelines, creating advisory committees and expanding research. Apart from addressing bias and discrimination (see Section 5.1), Japan and Singapore's AI guidelines also emphasise human dignity [79] and human centricity [80] and the Japanese government also created an Advisory Board on AI and Human Society to ensure the sustainability, prosperity and inclusivity from AI usage [102]. China intends to develop laws and ethical norms for AI governance [103], South Korea intends to create an ethics charter to provide ethical guidelines for AI and a "public-private partnership council" to examine technological risks [81], whereas Singapore's recently established Fairness, Ethics, Accountability and Transparency Committee will issue AI guidelines and codes of conduct to address bias and other ethical issues [104].

In 2017, the government of Germany released the world's first ethical rules for AVs. The rules highlight that human life should be prioritised above all others, echoing deontological principles, but also advocate utilitarian principles of damage minimisation so long as individuals are not discriminated based on personal characteristics and AVs are not programmed to "unconditionally save the driver" [105]. However, excluding personal characteristics may undermine the algorithm's ability to minimise damages during unavoidable accidents [106] and it is unclear how the two conflicting goals can be reconciled. The guidelines also assert that decisions in unavoidable accidents should not be programmed and that an independent agency should "systematically process the lessons learned" from these crashes [105]. Lastly, the guidelines highlight that the programming of AVs should be disclosed to be public and standards should be developed for "self-learning" processes, which should not be utilised for "safety-critical functions" unless deemed to satisfy safety requirements [103].

*5.3. Perverse Incentives*

Stakeholders in the AV ecosystem are motivated by different incentives in designing and operating algorithms in AVs: AV manufacturers profit from the sale of AVs; AV consumers purchase the AV and its services; Software and hardware companies that design the AVs' algorithms and hardware, ride-sharing and transportation network companies that sell AV services and data aggregators that use the data collected by AVs are motivated by different profit incentives [107]. This Section highlights how different incentives motivate stakeholders' design of AV algorithms in ways that perpetuate discrimination and new safety risks.

Firstly, manufacturers can program the AVs' algorithms to maximise profit rather than to ensure safe driving outcomes, which can yield unfair risk allocations that discriminate against other groups of road users. As the incentives of the manufacturer (to maximise sales and profits) and of the customer (to ensure their safety) are aligned, AV manufacturers can program AV algorithms to disproportionately allocate safety risks away from the AV passenger to other third parties [31]. Improving the safety of some individuals at another's "expense" may be considered unethical, even if the overall number of fatalities declines [33]. In a more extreme scenario, given existing product liability frameworks such as in the US that determine liability damages based on the amount of income "lost to dependents" [35], AV manufacturers could seek to maintain liability claims at a constant level by programming the AV's driving behaviour as a function of average income in a given district, implying that AVs would take more safety precautions in more "affluent" districts [35]. Doing so could appear discriminatory along income levels as it transfers safety risks from areas characterised by higher-income levels to areas characterised by lower-income levels.

Economic incentives of AV purchasers may also motivate manufacturers to design algorithms that discriminate between passengers of publicly-owned AVs and those of privately-owned AVs. Unlike publicly-owned AVs, private AV purchasers may expect their AVs to prioritise their safety over all others [89,92] and thus, AV manufacturers could program their algorithms according to these customer preferences to maximise profits. While doing so ensures AVs' commercial success, as it has been shown that consumers do prefer riding in AVs that prioritise passenger safety [108,109], safety risks would be transferred from users of privately-owned AVs, who benefit from their use, to individuals who cannot afford to purchase AVs. If private ownership of AVs is associated with lower allocated risks to its passengers, other group characteristics associated with private ownership, such as income levels, will be associated with lower allocated risks relative to that in publicly-owned AVs. This implies a potential for discriminating along income levels as the individuals who can afford to purchase AVs (and are therefore more shielded from safety risks) are likely to have higher average incomes than users of publicly-owned AVs.

In addition, profit incentives for product differentiation may lead to heterogeneity in algorithms' decision-making preferences, reduced traffic coordination among multiple AVs and new safety risks.

Occasionally, the AV may have to make assumptions about the behaviour of road users—assumptions that are shaped by its programmed rules and preferences, which may differ based on their manufacturer's configuration of different safety optima [35]. This lack of standardisation among AVs' driving styles may create a mismatch in expectations among road users and generate unexpected dynamics [110] that consequently increase the risk of collisions. These safety risks stemming from the interaction of AVs with heterogeneous decision-making preferences may imply a need for greater data sharing and standardisation of algorithms among AV developers, but concerns over privacy and intellectual property rights can impede firms' willingness to share data and these issues have yet to receive sufficient attention [35,110].

Lastly, various stakeholders in the AV value-chain that influence the design and operation of AV systems are motivated by different incentives that can potentially create negative externalities if not sufficiently coordinated. While AV manufacturers are motivated by profit measures such as the number of vehicles sold or their average margin per vehicle sold, ride-sharing and transportation network companies measure profits based on the number of trips completed, average km/mi travelled per trip, or average margin per km/mi travelled, whereas data aggregators maximise profits based on mapping the mobility data to online behavioural data and the value of the insights they can derive from the data for themselves and their customers [107]. Studies have yet to explore the implications of potentially misaligned incentives between AV stakeholders for AV safety, but some have noted that misaligned incentives severely impede the development of safe AI systems [111].

In summary, bias in the data, choice of variables and the algorithm's model can skew the AV's risk allocation decision that results in discrimination against certain groups of road users, and this is exacerbated by perverse incentives of AV stakeholders that benefit from these discriminatory outcomes. We also examined the role of ethics in AVs' decision-making—Ethical rules can be formalised from thought experiments, derived from well-established ethical theories and programmed in a top-down fashion, or can be learnt in a bottom-up fashion by ML algorithms, but each of these approaches contains its own set of limitations that can undermine safety and/or perpetuate discrimination (See Table 3 for the summary of the ethical issues below).

**Table 3.** Summary of the Ethical issues.

| | **Ethical Issues** | **Proposed Solutions/Steps Taken** |
|---|---|---|
| Bias | *Sources of bias in AV algorithms*<br>- Statistical bias and including personal characteristics in the data.<br>- Manufacturers and programmers can program algorithms to favour AV users' safety to boost profits.<br>- Large-scale replication of algorithmic preferences in AVs can perpetuate systemic discrimination.<br><br>*Challenges of detecting and correcting bias:*<br>- Algorithmic opacity masks decision-making logic.<br>- Data-driven and unpredictable nature of ML-based decisions makes it difficult to predict bias.<br>- Humans are excessively trusting of algorithmic decisions due to "automation bias".<br>- Difficult to prove discriminatory intent in algorithms. | *Proposed solutions*<br>- Modify the data, algorithm and output to offset bias.<br>- Measure and test for data bias, and identify the affected individuals.<br>- Clarify the standards to evaluate bias in algorithms.<br>- Increase transparency via traceability and interpretability.<br><br>*Steps taken*<br>- AI guidelines that emphasise on fairness, transparency and accountability—Japan, Singapore.<br>- Creating design and testing methods to mitigate bias and discrimination from AI—South Korea, UK.<br>- Prohibiting the use of sensitive personal data in automated decisions and mandating a right to explanation—EU GDPR. |

**Table 3.** *Cont.*

| | Ethical Issues | Proposed Solutions/Steps Taken |
| --- | --- | --- |
| Ethics | *Thought experiments—The Trolley Problem*<br>- Assumptions do not hold in actual driving scenarios.<br>- Participants' ethical reasoning may be inconsistent.<br>- Aggregating single trolley scenarios may create discriminatory patterns.<br>*Top-down approach*<br>*(i) Utilitarianism*<br>- Collective harm minimisation can penalise certain groups of individuals more than others.<br>- Risk of computational errors and issues with mathematically defining algorithmic preferences.<br>*(ii) Deontology*<br>- Rule conflicts, failure to cover all driving scenarios, and path dependencies create safety risks.<br>- Difficult to explicitly program ambiguous rules.<br>*Bottom-up approach*<br>- Difficult to specify a self-learning system's higher level goals and to ensure it expands its set of choices.<br>- The system can override its ethical rules.<br>- The system's decision-making is more opaque. | *Proposed solutions*<br>- Combine utilitarian and deontological ethics in the top-down approach.<br>*Steps taken*<br>- AI guidelines and committees to examine discrimination—Japan, Singapore, China, South Korea.<br>- Ethical rules for AVs—Germany: Promotes utilitarian ethics of damage minimisation but prohibits discrimination based on personal characteristics; Decisions for unavoidable accidents should not be programmed but independently assessed; Standards for self-learning processes should be developed and AVs' programming should be disclosed to the public; Prohibits use of self-learning systems for safety-critical functions unless proven sufficiently reliable. |
| Perverse incentives | - Manufacturers can design algorithms to favour passenger safety and tailor driving behaviour based on district affluence to reduce liability claims.<br>- Manufacturers' differentiation of algorithms can reduce road coordination and create safety risks.<br>- The incentives of other AV stakeholders in the supply chain can interact systemically and create safety risks. | *Proposed solutions*<br>- Greater data sharing, collaboration and standardisation of algorithms among AV developers to improve coordination of AVs on roads. |

## 6. Technical Concerns from Algorithmic Decision-Making in AVs

This Section discusses existing issues in the AV system's perception (Section 6.1), decision-making (Section 6.2) and control (Section 6.3) components, the limitations of existing AV safety verification and testing methods (Section 6.4), their implications for AV safety and the steps taken to manage these issues. Our study narrows the focus to issues in the AV's software components and issues specific to AI algorithms. Other technical issues that need to be addressed are cybersecurity risks, such as the vulnerability of AV sensors, communication networks and electric vehicle charging networks to denial of service attacks and data manipulation [112] (see Reference [45] for a detailed analysis of cybersecurity concerns in AVs).

*6.1. Perception*

Sensors in the AV's perception component are crucial for vehicle localisation but are limited by their inaccuracies and high costs. Global Navigation Satellite System (GNSS)-based sensors are costly and are still inaccurate and highly sensitive in urban environments [113]. For instance, GNSS-based sensors that integrate Global Positioning System (GPS) are still susceptible to the latter's localisation inaccuracies [114]. GNSS-based lane-level self-localisation methods are also susceptible to "multi-path interference" that occurs when the GPS signal is obstructed by external objects [113], satellite clock errors and inevitable inconsistencies between GNSS coordinates and High Definition (HD) Map coordinates [115]. Alternatively, visual sensors are less costly but are still inaccurate in adverse weather conditions and busy backgrounds, as they have been designed to operate on more clear images and videos [116–118]. Information from HD maps can be used to improve the images provided by visual sensors, but constructing HD maps requires large amounts of software and manual effort [114]. Lastly, Light Detection and Ranging (LiDAR) sensors are costly and, given their existing limitations in recognising "non-grounded objects", it is unclear if they can identify humans when the latter move unexpectedly [118]. Scholars recommend fusing different sensor types that have overlapping capabilities to reduce costs, achieve redundancy, and boost safety and performance [22,114].

In ML-based perception systems, such as neural networks, sensor inputs are susceptible to manipulation through adversarial samples, which are created by modifying camera images to provoke certain behaviours from the neural network, such as reducing the system's confidence on a prediction or causing it to misclassify inputs [119–121]. AV sensor inputs can be manipulated, such as slightly modifying road signs to cause the AV's neural networks to misclassify these signs, display erroneous behaviour and create road safety hazards [120]. To increase their resistance to manipulation, scholars propose ways to protect neural networks against adversarial samples. Papernot et al. [122] introduced defensive distillation that smooths the neural network models' gradients to reduce output variations around slight changes to inputs, but this method was later proven ineffective against stronger adversarial attacks [123]. Alternatively, Raghunathan et al. [123] proposed an approach to minimise the upper bound of the error induced from adversarial attacks and to produce a certificate for it [123], whereas Huang et al. [121] proposed automating the verification of the safety of ML-classified decisions.

Safety risks can also arise when errors in AV perception propagate to subsequent software components [22]. Perception algorithms process sensor inputs and generate outputs on their understanding of the AV's environment, where the latter could be inaccurate and used as inputs in decision-making algorithms that subsequently shape the AV's motor commands, potentially yielding unsafe driving behaviour. Such propagation of errors from the perception component contributed to Tesla's fatal AV accident in 2016 [77]. To ensure that the system accounts for uncertainty introduced at the perception component, estimating and minimising uncertainty in each individual component is critical, such as by using a Bayesian probability framework and Monte Carlo dropout sampling to estimate confidence scores of predictions generated by the AV's perception system [124,125]. Uncertainties in each component should also be communicated and well-integrated across all software components to provide an overall measure of the system's uncertainty to facilitate decision-making [77].

The AV system's performance can also be constrained by computationally demanding perception algorithms. Lin et al. [97] showed that object detection, tracking and localisation together account for over 94% of the AV system's computational power. These computational constraints hinder further improvements in accuracy that could be attained from adopting higher resolution cameras and using computational platforms such as graphics processor units (GPUs) to overcome these limitations will generate additional heat, significantly raising power consumption and reducing the AV's driving range and fuel efficiency [97]. While ML techniques such as Deep Neural Network architectures can improve object detection tasks, such as bounding-box detection that maximises the likelihood of detecting an object inside a box and semantic segmentation that classifies each pixel in the image space [22], they can introduce time delays when used to classify high-resolution images in real-time [22].

*6.2. Decision-Making*

AVs face several challenges in decision-making under dynamic road environments that are fraught with uncertainties and unpredictable movements of objects, such as road closures, accident clean-ups and other road users [37,126]. A major challenge is the potential failure to correctly interpret the meaning of certain decision-making rules during complex driving scenarios. The meaning of decision-making rules, such as ethical rules and traffic rules, may vary under different driving situations and usually requires human judgement or discernment [36]. Interpretational failures will instead lead the AV to display behaviours that contradict the principles that it was programmed with and potentially create safety risks to all road users.

Modelling and understanding human-vehicle interactions is essential for safe navigation in mixed traffic, to build consumer trust in AVs and promote their widespread adoption to realise their full safety benefits [45,127], but this remains a challenge for decision-making algorithms. Firstly, understanding humans in the AV is crucial to ascertain whether the human is prepared to regain control of the vehicle [128]. For instance, safety risks can emerge from handing over control to passengers if the AV system fails to recognise behavioural traits indicative of exhaustion or distraction. Secondly, understanding the intent of humans neighbouring the AV and in other vehicles is key for safe navigation. For instance, humans typically use hand gestures and other social cues to indicate their intention of breaking some traffic rules to facilitate traffic flow [36], or some nearby pedestrians may appear inattentive to the AV's movements. AVs will also have to negotiate with other road users during activities (e.g., giving way, overtaking, and merging), and this requires balancing between uncertainties in human behaviour while avoiding overly-defensive driving behaviour to ensure smooth traffic flows [126]. However, AVs may fail to correctly interpret or perform social cues, which can hinder other road users' ability to anticipate the AV's actions [36,129] and create mismatched expectations that can lead to accidents, which has already contributed to most of the collisions that have occurred during AV trials to date. For instance, accidents occurred when AVs stopped unexpectedly when human drivers do not expect them to stop, such as at an amber traffic light or at a congested intersection [129]. Despite these issues, limited studies have explored these human-vehicle interactions for AV vision and learning capabilities, which is crucial to address the aforementioned concerns [128].

Decision-making algorithms are also constrained by computational complexity and algorithms in other software components that can undermine the AV's performance and safety in dynamic environments. While studies have developed successful motion planning algorithms such as graph searches and rapidly-exploring random trees for AVs and other mobile robots, finding the optimal path is computationally expensive and not always possible [130,131]. In addition, amid multiple dynamic obstacles such as pedestrians and other road users, computationally demanding perception algorithms reduce the time available for motion planning algorithms to continuously compute new collision-free trajectories [50]. Motion planning algorithms also need to be well-integrated with and properly account for constraints faced by control algorithms such as the evolution of time, velocity and acceleration limits, to plan trajectories, but doing so requires more computational resources than existing processors can handle [37,132]. In this light, some researchers have begun developing path-planning algorithms that can account for perception uncertainties and control constraints to mitigate potentially dangerous scenarios [50]. In addition, vehicle trajectories that were initially considered safe can become dangerous upon unexpected environmental changes, such as when a moving obstacle obstructs the AV's view and perception of the initially planned trajectory [37]. New methods for incremental planning and adjustments of the plans can enhance AVs' adaptability to unexpected scenarios [37]. This could be supplemented with 5G networks to support nearly instantaneous decision-making by providing AVs with more information on nearby obstacles through more reliable vehicle-to-infrastructure (V2X) channels and at much faster speeds than existing 4G networks [133–135]. More research into the issues around shared AV service provision is required, where decision-making algorithms require sufficient computing capacity in exploring large decision spaces and catering to "spatiotemporally distributed" bookings in real-time [22].

*6.3. Control*

Control algorithms and their underlying models of vehicle motion have been developed with considerable success for trajectory tracking, which ensures that the AV moves along the path determined by its decision-making algorithms [136,137]. Many studies refer to "control algorithms" as "controllers" or "control strategies" [22,137,138]. However, safety risks can arise from control algorithms' potential inaccuracies in modelling the AV's motion, particularly amid unexpected road conditions.

Geometric and kinematic control algorithms are recognised for their simplicity and relatively low computational cost [139], but as they only model the vehicle's geometrical dimensions and kinematic properties such as acceleration and velocity [138], they can lead to inaccuracies and vehicle instability due to their neglect of vehicle dynamics. Without considering vehicle dynamics such as friction forces, tire slips and energy, geometric and kinematic control algorithms can lead to risky driving behaviour at high speeds where dynamics significantly influence the vehicle's motion, such as during sudden lane changes or attempts to avoid unexpected obstacles [138,140]. In the application of the "Pure Pursuit" geometric algorithm, where the vehicle is "in constant pursuit of a virtual moving point" [138], "rapid changes" in the vehicle's path during high-speed driving can cause the algorithm to "overestimate" the system's ability to produce steering inputs to correct the vehicle's movement, resulting in excessive steering and skidding of the rear vehicle [51,138]. Furthermore, setting control parameters to "compensate" for the neglect of dynamics renders geometric and kinematic control algorithms highly sensitive to parameter variations [139]. For instance, it is challenging to tune an optimal value of the "look-ahead distance" for the Pure Pursuit algorithm, which is measured from the vehicle's chosen "path point" from the vehicle's existing position [141], as values that are too large cause the vehicle to "cut corners" during sharp turns by deviating from the actual curved path, whereas values that are too small worsen the oscillation of the trajectory [51,141].

Dynamic, adaptive or model-predictive control algorithms have also been used in AVs but remain inaccurate when assumptions are violated and are computationally expensive. Firstly, dynamic control algorithms incorporate linear or nonlinear models of the vehicle's dynamics, mainly from tire forces which arise from the friction produced between the tire and road surfaces and are the main external influence of the vehicle's motion [51]. Linear models become inaccurate when the steering angle and lateral slip angle exceed five degrees, whereas nonlinear models are more accurate, particularly at high speeds and large steering angles, but are more computationally expensive [51,137]. Similar to geometric pure-pursuit algorithms, some dynamic control algorithms are still highly sensitive to variations of the "look-ahead distance" and unknown vehicle parameters such as tire-road frictions, whose values cannot be constantly obtained in real-time due to the high costs of installing additional sensors [51,142]. Secondly, adaptive control algorithms use ML techniques to tune the algorithm's parameters and thus are more robust to environmental changes, but they may be less efficient as ML requires processing vast amounts of data [51]. Thus, Amer et al. [51] recommends developing an adaptive geometric controller that has both the advantages of low computational cost afforded by geometric controllers and adaptive controllers' robustness to varying road conditions. Lastly, model-predictive control (MPC) algorithms account for system constraints, inputs and outputs to optimise actuator inputs and have been successfully used for AV trajectory tracking while meeting safety and time constraints. However, MPC requires highly complex and computationally demanding online optimisation, particularly when accounting for nonlinear vehicle dynamics [33,135,136,141]. Studies have proposed linearising non-linear vehicle models and relaxing some collision avoidance constraints to reduce these computational demands, which can be further facilitated in future with recent improvements in computational power and new developments of highly efficient algorithms for implementing MPC controllers in AVs [33,135,136]

Furthermore, all control algorithms face constraints induced by other software components, challenges in handling unexpected situations and also lack sufficient real-world testing. Firstly, most of the proposed control algorithms perform well only if the trajectories computed by decision-making algorithms (motion planners) are continuous [51] and have yet to account for time delays propagated by computationally demanding sensors that can significantly undermine vehicle stability [139]. Studies also show that trajectory tracking and vehicle stability can still be undermined during unexpected situations, such as emergency collision avoidance, when sudden path changes cause the AV's tires to sideslip [143]. Tires become highly saturated during sudden path changes, which require "large actuator inputs" within a limited timeframe [143]. Lastly, due to their large computational costs, most control algorithms have been tested only in simulations rather than in actual AVs [37,51,144] and they have only been validated under conditions of minimal parameter variations and unexpected environmental changes [138]. To ensure controllers' applicability to real-world AV implementations, studies have proposed using "hardware-in-the-loop" (HIL) simulations that includes a physical actuator in the simulation tests, developing a V2X system that utilises environmental data to update the AV's control parameters as driving conditions change and that is robust to wireless network disruptions, and developing controllers that integrate the steering, braking, and suspension controls during various road conditions [51,138].

*6.4. Safety Verification and Testing*

Existing AV testing methods contain many limitations that make them insufficient in demonstrating AV safety prior to deployment. Firstly, many developers conduct extensive road testing and analyse data such as the number of km/mi travelled, injuries, and fatalities to improve the AV's performance until a relatively low proportion of fatalities and injuries is achieved, but this requires AVs to be driven a significant number of miles that could take a long time [145,146]. Secondly, existing standards for systems safety requirements were designed for traditional system engineering processes where requirements are "known" and "unambiguously specified" (e.g., ISO Standard 26262) and involves first creating the functional requirements, annotating the safety-relevant requirements, allocating them to safety-critical subsystems and designing the latter according to these requirements [147], but this is incompatible with adaptive systems in AVs that learn from new data in real-time rather than just relying on clearly defined requirements [127,147]. Thus, different approaches are required to articulate safety requirements in AV systems.

Validating AV systems is also challenging due to the non-deterministic nature of their algorithms and the adaptive nature of ML-systems. Firstly, it is difficult to evaluate whether AV test results are correct as non-deterministic algorithms in AV systems produce non-repeatable and probabilistic outputs, which suggests potential differences in system behaviour under almost identical tests, its high sensitivity to minor changes in environmental conditions and potential differences in behaviour during real-life deployment and during testing and certification [127]. This necessitates a new testing approach that focuses on building sufficient confidence that the system displays the desired behaviour instead of expecting precise and unique outputs from certain inputs [127,147]. Secondly, the training data in ML-systems can contain accidental correlations that lead to erroneous predictions (overfitting), which must be detected and mitigated to prevent any drastic changes in the rules learnt by the system, but this remains a challenge and requires the use of expensive manually-labelled data [22,147].

ML algorithms are also prone to exhibiting erroneous corner-case behaviours that have already led to fatal accidents in AV trials, such as that of Tesla and Google [26]. However, existing means of detecting and correcting these behaviours in advance during testing remain highly dependent on the manual collection of labelled test data which is costly and difficult to scale [148]. Simulating corner-cases is less costly than testing, particularly for the AV's radar systems [149], but there are risks of biases and overfitting on simulated data, not all kinds of driving scenarios may be covered as even experienced test designers have blind spots, and the high sensitivity of non-deterministic systems to slight input changes exacerbates the challenge of formulating particular situations with particular combinations of inputs for the system to detect a corner-case [147]. Lastly, detecting corner-cases in ML-based software is more challenging than detecting bugs in traditional software, as the latter's logic is represented by easily examinable control flow statements but the logic in ML algorithms is learnt from data and embedded in highly nonlinear optimisation functions, which makes it more challenging to identify the inputs that trigger corner-case behaviours [150].

The limitations of existing testing and safety verification methods for AVs and ML can be improved in multiple ways, such as through fault injection, which is a widely recognised tool used for assessing safety and validating corner-cases in fault-tolerant mechanisms in autonomous systems [151], such as by randomly modifying the weights of neural networks and simulating erroneous inputs for sensors and maps to find defects in AV software that might be activated in unexpected scenarios [147,151,152]. Synthesis approaches and formal verification tools are popular means to verify AV control systems but are limited in deployment due to their high computational costs. Formal verification tools such as online verification of control algorithms and networks require traffic scenarios and road users to be represented in terms of probabilities, whereas traditional verification tools face challenges in modelling complex environments and specifying the system's desired properties [22,153]. Tian et al. [150] also highlight the need to make these verification methods more scalable for larger ML algorithms in real-world applications.

In summary, computational cost is a common issue in many algorithms across AV software components, modelling and understanding human-vehicle interactions remain key challenges for decision-making algorithms, and the non-deterministic and adaptive nature of ML-systems render existing testing and safety verification methods insufficient to ensure AV safety (See Table 4 for a summary of the technical issues).

**Table 4.** Summary of the technical issues.

| | Technical Issues | Proposed Solutions/Steps Taken |
|---|---|---|
| Perception | • GNSS sensors are costly and inaccurate when combined with GPS; LiDAR sensors are costly and struggle capturing unexpected movements; visual sensors are inaccurate in adverse weather.<br>• Sensor accuracy can be limited by object detection and tracking algorithms' computational demands.<br>• ML-based perception systems are susceptible to adversarial samples that force algorithms to misclassify images.<br>• Errors in the perception component can propagate across the system and undermine decision-making accuracy. | • Sensor-fusion [22,114].<br>• Methods to detect adversarial samples [122], minimise error deviations from attacks [123] and automate the verification of ML-decisions [121].<br>• Minimise uncertainty in individual software components, e.g., Bayesian probability framework, Monte Carlo dropout sampling [124,125].<br>• Communicate uncertainties across all software components and estimate overall system uncertainty [77]. |
| Decision-making | • Inaccurate interpretations of ethical and traffic rules and human interactions create mismatched driving expectations.<br>• Constrained by delays propagated by perception algorithms.<br>• Planned trajectories may not account for unexpected obstacles.<br>• Not accounting for control constraints can create discrepancies between planned and executed trajectories.<br>• Insufficient computing capacity to meet the demands of shared AV services within time constraints. | • Expand research on modelling human-vehicle interactions [126,128].<br>• Incremental planning and adjustments of planned trajectories are required to incorporate unexpected environmental changes [37].<br>• Use data from 5G network channels to adapt to unexpected changes at higher speed and reliability [133–135]. |
| Control | • Geometric and kinematic controllers neglect vehicle dynamics, potentially resulting in excessive skidding during sudden path changes and high sensitivity to parameter changes.<br>• Dynamic controllers based on linear models remain inaccurate in variable conditions and nonlinear models are more complex.<br>• Adaptive and model-predictive controllers are more robust to environmental changes but more computationally costly.<br>• Most controllers lack real-world testing, are not prepared for unexpected parameter and environmental changes, highly sensitive to model assumptions and sensor-induced delays, and inaccurate during sudden path changes. | • Develop adaptive geometric controllers to reduce computational complexity [51].<br>• Linearise non-linear vehicle models and relax some model constraints to reduce MPCs' computational costs [37,137,138].<br>• Hardware-in-the-loop simulation to better reflect actual driving [51].<br>• Use data from a V2X system to update control parameters in real-time [138].<br>• Integrate steering, braking and suspension controls [51]. |

Table 4. *Cont.*

| | Technical Issues | Proposed Solutions/Steps Taken |
|---|---|---|
| Testing and verification | • Statistically proving AV safety via road testing is impractical.<br>• Current safety specification standards that assume deterministic system properties are inapplicable to ML-systems.<br>• Challenges in validating AV safety:<br>  (1) Outputs of non-deterministic algorithms can differ in almost identical test scenarios.<br>  (2) Erroneous correlations in the data are difficult to detect and costly to correct by acquiring test data.<br>• Challenges in detecting corner-cases:<br>  (1) Manual data collection is costly and difficult to scale.<br>  (2) Simulated data on corner-cases can be biased and may not cover all cases.<br>  (3) Algorithmic opacity makes it difficult to identify inputs that caused certain outputs. | • Improve testing and safety verification methods for AVs and ML algorithms:<br>  (1) Fault injection to identify unexpected defects, e.g., modifying neural networks, feeding erroneous inputs to sensors [147,151,152].<br>  (2) Formal verification tools to verify controllers and networks [22,153].<br>  (3) Improve the scalability of verification methods for larger ML algorithms [150]. |

## 7. Discussion

This study explored algorithmic decision-making in AVs and identified several key issues in the design and operation of AV algorithms that can create new safety risks and potentially lead to discrimination. We also explored various steps taken to address these issues, ranging from regulation to address bias and discrimination, issuing voluntary ethical guidelines for AI systems and AVs, to technical tools for improved AV testing and verification. More can be done to address algorithmic bias specific to AVs, trade-offs between ethical rules, perverse incentives of AV stakeholders and various technological issues in the AV system.

Firstly, bias may be introduced into the AVs' algorithms unintentionally or intentionally by AV manufacturers and programmers through the data, model and use of sensitive variables, which causes the algorithms to allocate more safety risks to certain types of road users. The opacity and unpredictability of ML algorithms, humans' excessive trust in algorithms and the seemingly objective nature of algorithmic decisions pose obstacles to detecting algorithmic bias, and the potential effects of systemic discrimination resulting from large-scale replication of algorithmic preferences across AV fleets remain underexplored. Existing steps taken to tackle bias and discrimination are not specific to AVs, such as creating advisory committees to examine AI-related risks, issuing AI guidelines, and expanding research to develop technical solutions. Scholars recommend supplementing these technical solutions with consensus building around the criteria and standards to evaluate bias in algorithms, such as the definition of "fairness" [30,154] to align algorithmic decisions in AVs and other AI systems with societal values. Doing so is key to ensuring that technological solutions such as AVs serve human needs and yield truly smart and sustainable cities.

The EU's GDPR aims at increasing transparency around algorithmic decision-making to tackle bias and discrimination by prohibiting the use of personal characteristics and mandating a right to an explanation for automated decisions, but several gaps in the GDPR's right to an explanation need to be addressed. Firstly, the explanations provided by ML algorithms may not fit the GDPR's legal conceptions of "meaningful information" as some explanations regarding the type of ML model, training data and model testing methods may not be directly relevant to particular individual data subjects or deemed "meaningful" by the inquirer of the explanation [47]. To "whom and when" information should be considered meaningful remains ambiguous [155]. Secondly, such explanations may not effectively convey the risks of algorithmic decision-making, as individuals are likely to lack the time, resources and expertise to comprehend these explanations [47]. Thirdly, explanations of individual aspects of ML decision-making processes (e.g., information regarding the "existence or logic" of a single decision) alone are not useful in identifying algorithmic discrimination, which would also require looking "across inputs and outputs" to examine the classification of other individuals or groups of individuals [156]. Thus, regulation needs to go beyond ex-ante or ex-post explanations to look across and continuously review the entire system's inputs, outputs and logic as algorithms are being updated with new data [156]. Algorithmic transparency should also target the type of decision being made and empower its observers to understand, challenge and improve the relations between humans and the algorithms across the system [47,157,158]. Furthermore, currently lacking in the development of smart cities is greater citizen engagement in designing and implementing technological solutions, which is critical to yield truly inclusive smart cities [27,159].

This study also analysed the approaches to construct and implement ethical rules into AVs' algorithms and identified several issues that can create new safety risks and potential discrimination. Firstly, while trolley problem experiments can provide insights into individual preferences and judgements that can inform the AV's decision-making, they make unrealistic assumptions, can produce unreliable experimental responses and potentially yield unintended systemic patterns resulting from aggregated individual trolley problems [31]. Secondly, programming ethical rules in AVs using the top-down and bottom-up approach introduces several issues. Under the top-down approach, programming utilitarian ethical principles into AVs can produce discriminatory outcomes as collective rather than individual harms are minimised, whereas deontological ethical rules can create safety risks due to potential rule conflicts, insufficient coverage of all driving scenarios and difficulties of translating legal ambiguities into programmable code [36,65,96]. The apparent dichotomy between utilitarian and deontological ethical rules highlights trade-offs between safety as a public good and as an individual good, raising further questions on when and how AVs, or AI systems in general, should prioritise individuals over society, and vice versa [154]. The bottom-up approach allows the AV to construct its own rules but can introduce new safety risks due to design constraints, the potential for the algorithm to override its programmed ethical rules in the learning process and increased opacity of decision-making. While enabling AVs' ethical reasoning to evolve dynamically as it accumulates driving experience, the bottom-up approach could be supplemented with top-down ethical principles that represent ideals against which the system's decision-making is evaluated [94,160,161].

Key stakeholders in the AV value-chain can be motivated by perverse incentives to design AV algorithms in ways to favour the safety of certain groups of road users over others, as discussed in Section 5.3, where manufacturers can prioritise the safety of AV passengers and particularly those in privately-owned AVs over other groups of road users to maximise profit or program the AV to drive more cautiously in high-income districts to minimise liability claims from accidents. Product differentiation among AV developers can result in heterogeneous algorithmic preferences, reduced traffic coordination and new road safety risks, which could suggest the need for standardising algorithmic preferences but doing so can possibly entail trade-offs for systemic discrimination if AVs display similar discriminatory behaviour at a large-scale. Other AV stakeholders such as software operators, data aggregators and ride-sharing companies are motivated by different profit incentives that can create systemic effects for safety and discrimination, which have not been explored in the literature. Given their key role in shaping "smart cities agendas and policies" [7], more action is required to align AV stakeholders' incentives with sustainability objectives.

Lastly, existing technical issues in AV software and hardware can create new safety risks and constrain further safety improvements for AVs. The inaccuracies of existing sensors can undermine AV perception during harsh weather conditions, the high computational demands of perception algorithms can constrain overall system performance, and ML perception algorithms' susceptibility to manipulation by adversarial samples and the lack of integration across all software components can cause perception errors to distort decision-making. Currently, decision-making algorithms still struggle navigating in dynamic environments due to potential interpretational failures and challenges in understanding human-vehicle interactions, as well as face additional computational constraints imposed by other software components and unexpected environmental changes. Furthermore, control algorithms' accuracy in tracking the AV's path is limited in various ways, ranging from the inaccuracies and potential vehicle instability stemming from geometric and kinematic control algorithms' neglect of vehicle dynamics, the computational cost of more complex dynamic, adaptive and model-predictive control algorithms, insufficient real-world testing of control algorithms and failures in accounting for time delays in other software components and handling unexpected driving situations. To date, it is difficult to verify AV safety prior to deployment as extensive test-driving is insufficient, existing safety specification standards are incompatible with ML-systems in AVs, and validating training data of ML-systems and detecting corner-cases before deployment remain open challenges. More research is required to address the high costs and low scalability of many system verification tools for AVs. It is also important to supplement these verification tools with adaptive policies and regulation that involves knowledge generation through investigative programmes, reviewing that knowledge such as through safety review boards and altering safety requirements in tandem with technological developments [41,145]. Other technical issues include the detection of inevitable sensor failures and cybersecurity risks.

These issues illustrate the broader concerns in the smart cities literature that technological smartness may not lead to truly smart and sustainable cities. The safety risks and discrimination that can arise from algorithmic decision-making in AVs can threaten public well-being, social equity, inclusivity, consumer trust in AVs and consequently, hamper the move to smarter and more sustainable cities. It is therefore critical for scholars, technology entrepreneurs and policymakers to consider these issues in the design, operation, and governance of AV algorithms. Instead of following smart mobility ideas only for their technological "smartness", these technological solutions must be supplemented with the smartness of policymakers, technocrats, urban leaders, and residents, as well as policy designs and action plans to achieve true urban smartness [5,53].

## 8. Conclusions

This study serves to deepen the understanding of algorithmic decision-making in AVs that have significant implications for smart and sustainable cities. Our first research question was addressed by analysing the most prominent concerns in the literature pertaining to algorithmic decision-making in AVs, namely algorithmic bias, ethical rules, perverse incentives, and technical issues stemming from different software components and testing procedures. By examining these ethical and technical issues, this study sheds light on the significant roles played by key actors in both the technological landscape and the wider AV ecosystem and their interactions in shaping the societal risks posed by AVs. These issues are not exhaustive and thus can be expanded in future work to include other significant factors influencing AVs' promise for smart and sustainable cities, particularly the barriers to consumer acceptance of AVs and the complementary role played by other technologies such as wireless charging and power grids as elements of the policy mix in advancing smart and sustainable mobility and improving environmental outcomes [15,66]. Secondly, we demonstrated the significance of these issues for smart and sustainable cities by examining the mechanisms and interactions between algorithmic processes and stakeholders in the AV landscape that can create new safety risks and discriminatory outcomes. We then addressed our third research question by examining the variety of proposed solutions to resolve these issues and their effectiveness.

The analysis conducted here reveals several research gaps that deserve further examination to advance AV's promise for smart and sustainable cities. Firstly, much of the discussion on algorithmic biases applies to all ML-systems in general, and thus more research is needed to analyse biases and their implications for safety in the context of AVs. Cross-disciplinary research and consensus-building regarding the definition and choice of benchmarks to evaluate bias are also required. Secondly, systemic discrimination from the aggregation of biased algorithmic preferences should be further analysed in the context of large-scale AV deployment. Thirdly, more research is required to analyse the trade-offs from using different types of ethical rules for AVs and from standardising AV algorithms. When and how individuals' safety should be sacrificed for (or prioritised over) public safety are key questions that need to be answered. Fourthly, future work should examine the implications of AV stakeholder's incentives, their interactions across the value-chain, and the ways to keep perverse incentives in check, particularly through greater accountability. Future work can expand the limited research on the types of accountability mechanisms for different AV stakeholders and the challenges of holding AV stakeholders accountable for their algorithms. Lastly, developing more accurate sensors, models of human-vehicle interactions, more efficient decision-making and control algorithms and new standards for verifying unpredictable ML-systems is crucial to enhance the safety and reliability of AV operations.

**Author Contributions:** H.S.M.L. and A.T. have contributed evenly to the design, analysis, writing of the manuscript. H.S.M.L. carried out most of the data collection and A.T. conceived the manuscript.

**Funding:** This research was funded by the Lee Kuan Yew School of Public Policy, National University of Singapore through the Start-up Research Grant.

**Acknowledgments:** Araz Taeihagh is grateful for the support provided by the Lee Kuan Yew School of Public Policy, National University of Singapore through the Start-up Research Grant.

**Conflicts of Interest:** The authors declare no conflicts of interest.